\documentclass[twocolumn,showpacs,aps,prd,floatfix]{revtex4}
\usepackage{graphicx}
\usepackage{dcolumn}
\usepackage{amsmath}
\usepackage{epsfig}
\usepackage{longtable}
 
\newcommand{\BaBarNumber}     {PUB-05/023}
\newcommand{\SLACPubNumber} {11333}
 
\input babarsym

\long\def\inst#1{\par\nobreak\kern 4pt\nobreak
    {\it #1}\par\vskip 10pt plus 3pt minus 3pt}
                                                                                                          
\begin{document}
 
\begin{flushleft}
\babar-\BaBarNumber \\
SLAC-PUB-\SLACPubNumber\\
\end{flushleft}
 
\title{ 
\Large \bf\boldmath 
Dalitz Plot Analysis of
   \boldmath{$D^0 \to \overline{K}{}^0 K^+ K^-$}
}

%
\author{B.~Aubert}
\author{R.~Barate}
\author{D.~Boutigny}
\author{F.~Couderc}
\author{Y.~Karyotakis}
\author{J.~P.~Lees}
\author{V.~Poireau}
\author{V.~Tisserand}
\author{A.~Zghiche}
\affiliation{Laboratoire de Physique des Particules, F-74941 Annecy-le-Vieux, France }
\author{E.~Grauges}
\affiliation{IFAE, Universitat Autonoma de Barcelona, E-08193 Bellaterra, Barcelona, Spain }
% ...
\collaboration{The \babar\ Collaboration}
\noaffiliation

\begin{abstract}
A Dalitz plot analysis of approximately 12500 $D^0$ events reconstructed 
in the
hadronic decay
$D^0 \to \overline{K}^0 K^+ K^-$ is presented. 
This analysis is based on a data sample of 91.5 \invfb collected with the 
\babar\ detector 
at the PEP-II asymmetric-energy $e^+ e^-$ storage rings at SLAC
running at center-of-mass energies on and 40 MeV below the \Y4S resonance.
The events are selected from  
$e^+ e^- \to c \bar c$ annihilations using the decay $D^{*+} \to D^0 \pi^+$.
The following ratio of branching fractions has been obtained: 
$$BR = \frac{\Gamma(D^0 \to \overline{K}{}^0 K^+ K^-)}{\Gamma(D^0 \to \overline{K}{}^0 \pi^+ \pi^-)} =(15.8 \pm 0.1 \;(\text{stat.})
\pm 0.5 \;(\text{syst.})) \times 10^{-2}.$$
Estimates of fractions and phases for resonant and 
non-resonant
contributions to the Dalitz plot are also presented.
The $a_0(980) \to \bar K K$ projection has been extracted with little 
background.
A search for CP asymmetries on the Dalitz plot has been performed.

\end{abstract}


\maketitle



\section{Introduction}\label{sec:intro}
The Dalitz plot analysis is the most complete method of studying
the dynamics of three-body charm decays.  
These decays are
expected to proceed through intermediate quasi-two-body modes 
\cite{two} and experimentally
this is the observed pattern.
Dalitz plot analyses can provide new information on the resonances that
contribute to observed three-body final states. 
  
In addition, since the intermediate quasi-two-body modes are dominated by
light quark meson resonances, new information on light meson spectroscopy 
can be obtained. Also,
old puzzles related to the parameters and the internal structure of several
light mesons can receive new experimental input. 

Puzzles still remain in light meson spectroscopy. There are new 
claims for the existence of broad states close to threshold
such as $\kappa(800)$ and $\sigma(500)$~\cite{e791}. 
The new evidence has reopened discussion of the composition of the ground state
$J^{PC}=0^{++}$ nonet, and of the possibility that states such as the 
$a_0(980)$
or $f_0(980)$ may be 4-quark states due to their proximity to the
$\bar K K$ threshold~\cite{q4}. This hypothesis can only be tested through 
an accurate measurement of branching fractions and couplings to  
different final states. In addition, comparison between the production
of these states in decays of differently flavored charmed 
mesons $D^0 (c \bar u)$, $D^+(c \bar d)$ and $D_s^+(c \bar s)$ 
can yield new information on their 
possible quark 
composition. Another benefit of studying charm decays 
is that, in some cases, partial wave analyses
are able to isolate the scalar contribution 
almost background free. 

This paper focuses on the study of the three-body 
$D^0$ meson decay 
$$D^0 \to \overline{K}{}^0 K^+ K^-,$$
where the $ \overline{K}{}^0$ is detected via the decay 
$K^0_S \to \pi^+ \pi^-$. All references in this 
paper to an explicit decay mode, unless otherwise
specified, imply the use of the charge conjugate decay also.

This paper is organized as follows. Section II  briefly describes the 
\babar\ detector, while Section III gives details on the event reconstruction.
Section IV is devoted to the evaluation of the efficiency and the 
measurement of the branching fraction is reported in Section V. Section VII
deals with a partial wave analysis of the $K^+K^-$ system, while sections
VI, VIII, IX and X describe the Dalitz plot analysis.

\section{The \babar\ Detector and Dataset} \label{sec:detector}

The data sample used in this analysis consists of 91.5 \invfb recorded 
with the \babar\ detector at the SLAC \pep2\ storage rings. 
The PEP-II facility operates nominally at the \Y4S 
resonance, 
providing collisions of 9.0 \gev electrons on 3.1 \gev positrons. The data 
set includes 
82~\invfb collected in this configuration
(on-resonance) and 9.6~\invfb collected at a c.m. energy 40 MeV below
the \Y4S resonance (off-resonance).

The following is a brief summary of the 
components important to this analysis. A more complete overview of the 
\babar\ detector can be found
elsewhere~\cite{ref:babar}.
The interaction point is surrounded by a five-layer 
double-sided silicon vertex tracker (SVT) and a 40-layer drift chamber (DCH) 
filled with a gas mixture of helium and isobutane, all within a 1.5-T 
superconducting 
solenoidal magnet. 
In addition to providing precise spatial hits for tracking, 
the SVT and DCH 
measure specific energy loss $dE/dx$, which provides
particle identification for low-momentum charged particles.
At higher momenta ($p>0.7$~\gevc) pions and kaons are identified by
Cherenkov radiation observed in the DIRC, a detector designed to 
measure Cherenkov angles of photons internally reflected in the radiator. 
The typical separation between pions and kaons varies from $8 \sigma$ 
at 2 \gevc  to 2.5$\sigma$ at 4 \gevc.

\section{Event Selection and $D^0$ Reconstruction}

This analysis includes a measurement of the branching ratio:
$$BR = \frac{\Gamma(D^0 \to \overline{K}{}^0 K^+ K^-)}{\Gamma(D^0 \to \overline{K}{}^0 \pi^+ \pi^-)}.$$
Therefore the selection of both the data samples corresponding to
$$D^0 \to \overline{K}{}^0 \pi^+ \pi^-, \qquad (1)$$
and
$$D^0 \to \overline{K}{}^0 K^+ K^- \qquad (2)$$
is described.
The two final states are referred to collectively as $K^0 h^+ h^-$.

The decay $D^{*+} \to D^0 \pi^+$ is used
to distinguish between $D^0$ and $\overline{D}{}^0$ and to reduce
background. For example, the Cabibbo-favoured decays under study are
$$D^{*+} \to \begin{array}[t]{l} D^0 \pi^+ \\ \to \overline{K}{}^0 \pi^+ \pi^- \end{array},$$
$$D^{*-} \to \begin{array}[t]{l} \overline{D}{}^0 \pi^- \\ \to K^0 \pi^+ \pi^- \end{array}.$$
The charge of the slow $\pi^{\pm}$ from $D^*$ decay
(referred to as the slow pion $\pi^+_S$)  identifies the flavor of the $D^0$ and $K^0$ 
(for the latter, ignoring the small contribution from doubly-Cabibbo-suppressed decay of the $D^0$).

A $D^0 \to K^0_S h^+ h^-$ candidate is reconstructed from 
a $K^0_S \to \pi^+ \pi^-$
candidate plus two additional charged tracks, each with at least 12 hits in the
DCH. The slow pion is required to have  
momentum less than 0.6 \gevc and to have
at least 6 hits in the SVT. In addition, all the tracks are 
required to have transverse momentum $p_T>$ 100 MeV/$c$ and, 
except for the $K^0_S$ decay pions,
to point back to the nominal collision axis
within 1.5 cm transverse to this axis and within $\pm$ 3 cm 
of the nominal interaction point along this axis.

A $K^0_S$ candidate is reconstructed by means of a vertex fit to a pair of 
oppositely-charged tracks with the $K^0$ mass constraint.
The reconstructed $K^0_S$ candidate is then fit to a common vertex with all 
remaining combinations of
pairs of oppositely-charged tracks to form a $D^0$ candidate vertex. 
$K^0_S$ candidates are further required to have a flight distance 
greater than 0.4 cm 
with respect to the candidate $D^0$ vertex. 
The $D^0$ candidate is then combined with each slow pion candidate, 
and fit to a common $D^*$ vertex, which is constrained to be located 
in the interaction region. In all cases the fit probability is required 
to be greater than 0.1\%.  

To reduce combinatorial background, a $D^0$ 
candidate is required to have a  
center of mass momentum  greater than 2.2 \gevc. 

Kaon identification is performed by combining
$dE/dx$ information from the
tracking detectors with associated Cherenkov angle and photon information 
from the
DIRC. 
The resulting  efficiency is
above 95\% for kaons with less than 3 GeV/$c$ momentum that reach the DIRC.

\begin{figure}
\begin{center}
\includegraphics[width=9cm]{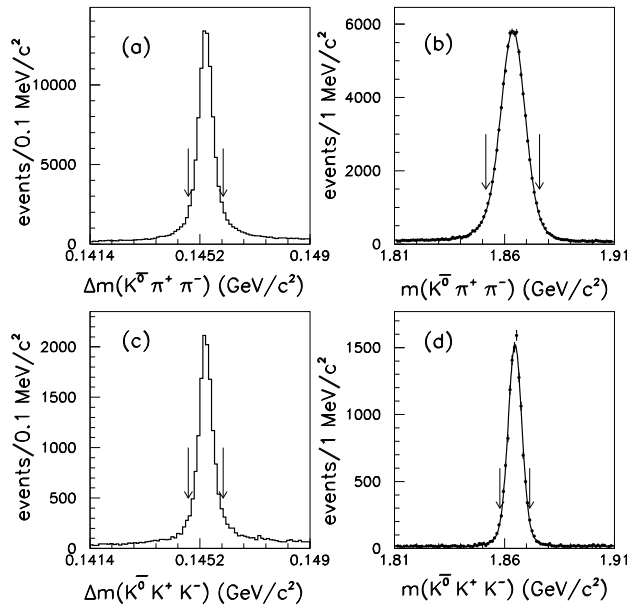}
\caption{a) and c) The $\Delta m$ distributions for 
$D^0 \to \overline{K}{}^0 h^+ h^-$ candidates,
for events in which the
$\overline{K}{}^0 h^+ h^-$  invariant mass is within two standard deviations of the $D^0$ mass value.
The arrows indicate 
the region of $\Delta m$ used to select the $D^0$ candidates.
b) and d) $\overline{K}{}^0 h^+ h^-$ mass distributions for events in which
$\Delta m$ is within 600 KeV/$c^2$ of the mean
$\Delta m$ value for signal events.
The arrows indicate 
the region of $m(\overline{K}{}^0 h^+ h^-)$ used to produce the $\Delta m$ distributions.
}
\label{fig:fig1}
\end{center}
\end{figure}

Each $D^0$ sample is characterized by the distributions of two
variables, the invariant mass of the candidate $D^0$, and the difference
in invariant mass of the $D^{*+}$ and $D^0$ candidates
$$\Delta m = m(\overline{K}{}^0 h^+ h^- \pi_S^+) - m(\overline{K}{}^0 h^+ h^-).$$
The distributions of $\Delta m$
for those candidates for which $m(\overline{K}{}^0 h^+ h^-)$ is within two standard 
deviations of the $D^0$ mass value
are shown in Fig.~\ref{fig:fig1}(a) and Fig.~\ref{fig:fig1}(c)
for $\overline{K}{}^0 \pi^+ \pi^-$ and  $\overline{K}{}^0 K^+ K^-$ respectively. 
Strong $D^*$ signals are apparent.
Fits to these
distributions produce consistent means and widths for the two
channels: $\Delta m = 145.41 \pm 0.01$ MeV/$c^2$, $\sigma=304 \pm 4$ KeV$/c^2$
in the case of decay channel $\overline{K}{}^0\pi^+\pi^-$ (statistical errors  only).

The $\overline{K}{}^0 h^+ h^-$ mass distributions for candidates that fall within 
$\pm$ 600 KeV$/c^2$ (two standard deviations) of the central value
 of the $\Delta m$ distribution are shown in Fig.~\ref{fig:fig1}(b) and 
 Fig.~\ref{fig:fig1}(d) for $\overline{K}{}^0 \pi^+ \pi^-$ and 
$\overline{K}{}^0 K^+ K^-$ respectively.

Fitting the $\overline{K}{}^0 h^+ h^-$ mass distributions using a 
linear background and a Gaussian function for the signal gives the following
mass and width (statistical errors only) for the decay $D^0 \to \overline{K}{}^0 \pi^+ \pi^-$:
\begin{center}
$m=1863.65 \pm 0.06$ MeV/$c^2$; $\sigma=6.10 \pm 0.02$ MeV/$c^2$
\end{center}
and for $D^0 \to \overline{K}{}^0 K^+ K^-$:
\begin{center}
$m=1864.74 \pm 0.03$ MeV/$c^2$; $\sigma = 3.37 \pm 0.03$ MeV/$c^2.$
\end{center} 
The mass resolution for reaction (2) is much better than that for reaction (1)
because of the much smaller Q-value involved (380 MeV/$c^2$ compared to 1088 MeV/$c^2$). The $\approx$ 1 MeV/$c^2$ shift between the two mass 
measurements is 
within the expected systematic error and is due to the different
kinematics of the two $D^0$ decay modes. 
\section{Efficiency}

The selection efficiency for each of the $D^0$ decay modes is
determined from a sample of Monte Carlo events in which each 
decay mode was generated
according to phase space (i.e. such that the Dalitz plot is uniformly
populated). These events were passed through a full detector
simulation based on the GEANT4 toolkit \cite{geant} and subjected to the same
reconstruction and event selection procedure as were the data. The distribution
of the selected events in each Dalitz plot is then used to
determine the relevant reconstruction efficiency. Typical Monte Carlo samples used to 
compute these efficiencies consist of 2 $\times 10^5$ generated events.
\begin{figure*}
\begin{center}
\includegraphics[width=12cm]{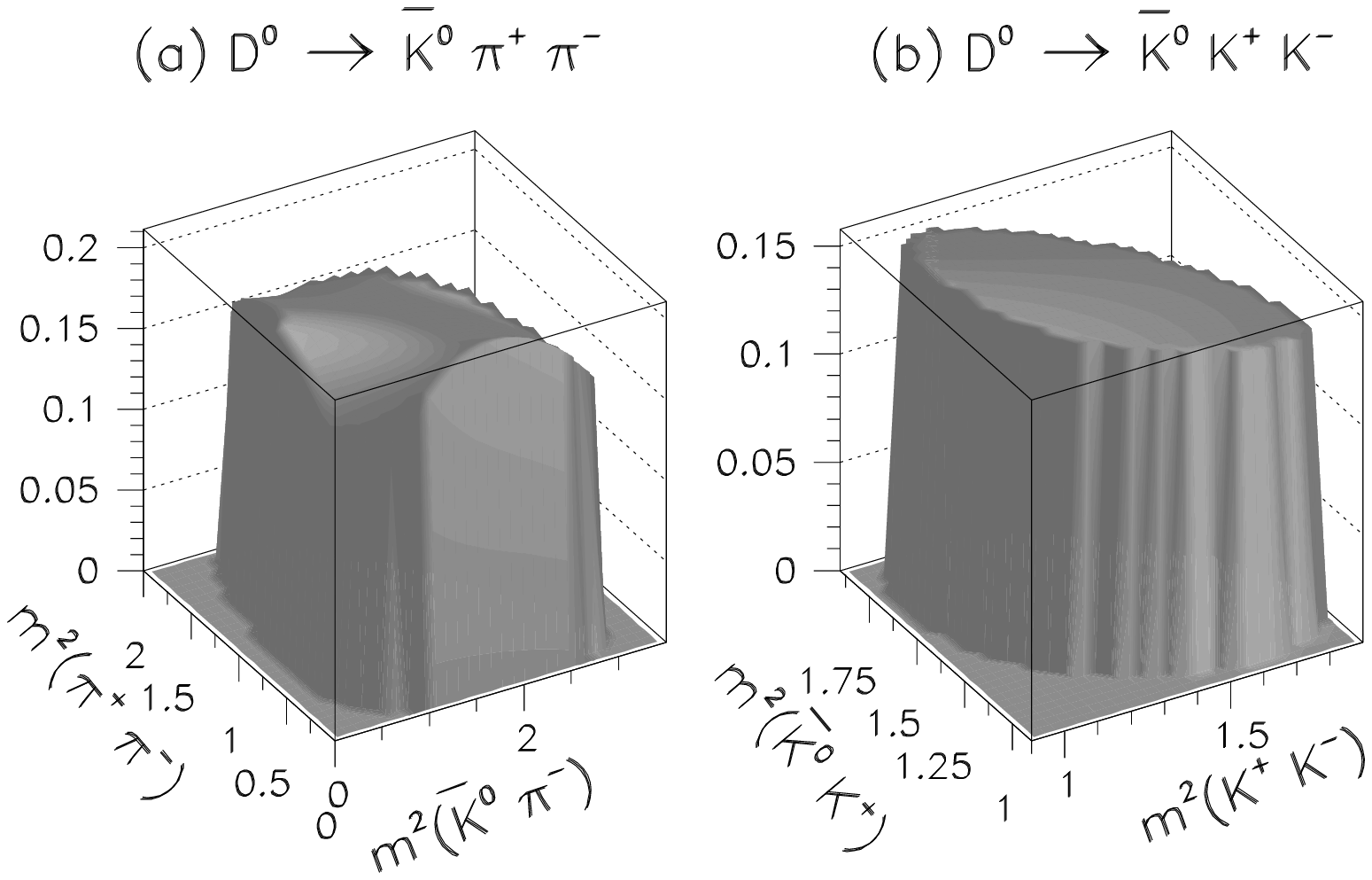}
\caption{Efficiency on the Dalitz plot for 
(a) $D^0 \to \overline{K}{}^0 \pi^+ \pi^-$ and 
(b) $D^0 \to \overline{K}{}^0 K^+ K^-$. 
} 
\label{fig:fig2}
\end{center}
\end{figure*}
Each Dalitz plot is divided into small cells and the efficiency 
distribution fit to
a third-order polynomial in two dimensions.
Cells with fewer than 100 generated events were ignored in the fit.
The resulting $\chi^2$ per degree of freedom ($\chi^2/NDF$)
is typically 1.1 using $\approx$ 500 cells. 
The fitted efficiencies are shown in Fig.~\ref{fig:fig2}. 
Using the weighting procedure described in the next 
section the weighted efficiencies values (17.94 $\pm$ 0.25)\% for 
$\overline{K}{}^0 \pi^+ \pi^-$ and (16.56 $\pm$ 0.38)\% 
for $\overline{K}{}^0 K^+ K^-$ are obtained. The above errors include the 
uncertainties on the weighting procedure. 
 
\section{Branching Fractions}

Since the two $K^0 h^+ h^-$ decay channels have similar topologies,
the ratio of branching fractions, calculated relative to the 
$\overline{K}{}^0\pi^+ \pi^-$ decay mode, is expected to have a reduced
systematic uncertainty.
This ratio is evaluated as
$$BR = \frac{\sum_{x,y} \frac{N_1(x,y)}{\epsilon_1(x,y)}}{\sum_{x,y}\frac{N_0(x,y)}{\epsilon_0(x
,y)}},$$
where $N_i(x,y)$ represents the number of events measured for channel $i$,
and $\epsilon_i(x,y)$ is the corresponding efficiency in a given Dalitz 
plot cell $(x,y)$ .

To obtain the yields and measure the relative branching fractions,
each $\overline{K}{}^0 h^+ h^-$ mass distribution 
is fit assuming a double 
Gaussian signal and linear background where all the parameters are floated, 
as shown in Fig.~\ref{fig:fig1}(b) and Fig.~\ref{fig:fig1}(d). 
The number of signal events is calculated as the difference between the
total number of events from the fit and 
the integrated linear background function in the same mass range.
The region used is within $\pm 6 \sigma$ of the $D^0$ mass.
Selecting events within three standard deviations of the central value
of the $\Delta m$ distribution,
the fits give the following yields for the two channels.
$$D^0 \to \overline{K}{}^0 \pi^+ \pi^-: \quad N=92935 \pm 305$$
$$D^0 \to \overline{K}{}^0 K^+ K^-: \quad N=13536 \pm 116.$$

Systematic errors take into account effects due to the use of 
selection regions for the $\Delta m$ 
distribution, the use of particle identification, the different fitting models 
used to subtract the background, $K^0_S$ reconstruction, and uncertainties in the 
calculation of the efficiency on the Dalitz plot due to Monte Carlo statistics. 
The resulting systematic error is dominated by the uncertainty due to 
efficiency correction on the Dalitz plot.

The resulting ratio is:
$$BR = \frac{\Gamma(D^0 \to \overline{K}{}^0 K^+ K^-)}{\Gamma(D^0 \to \overline{K}{}^0 \pi^+ \pi^-)} =$$
$$  (15.8 \pm 0.1 \;(\text{stat.}) \pm 0.5 \;(\text{syst.})) \times 10^{-2}$$
to be compared with the PDG value of: $(17.2 \pm 1.4) \times 10^{-2}$~\cite{pdg}. 
The best previous measurement of this branching fraction comes from the CLEO 
experiment (136 events for reaction (2)), which obtains the value
 $BR=(17.0 \pm 2.2) \times 10^{-2}$~\cite{cleo2}.

The branching ratio measurements have been validated using a 
fully inclusive
$e^+ e^- \to c \bar c$ Monte Carlo simulation incorporating all known 
$D^0$  decay modes.
The Monte Carlo events were subjected
to the same reconstruction, event selection and analysis procedures as for the 
data. The results were found to be consistent, within statistical uncertainty, 
with the branching fraction values used 
in the Monte Carlo generation.

\section{Dalitz Plot for $D^0 \to \overline{K}{}^0 K^+ K^-$}

\begin{figure}
\begin{center}
\includegraphics[width=10cm]{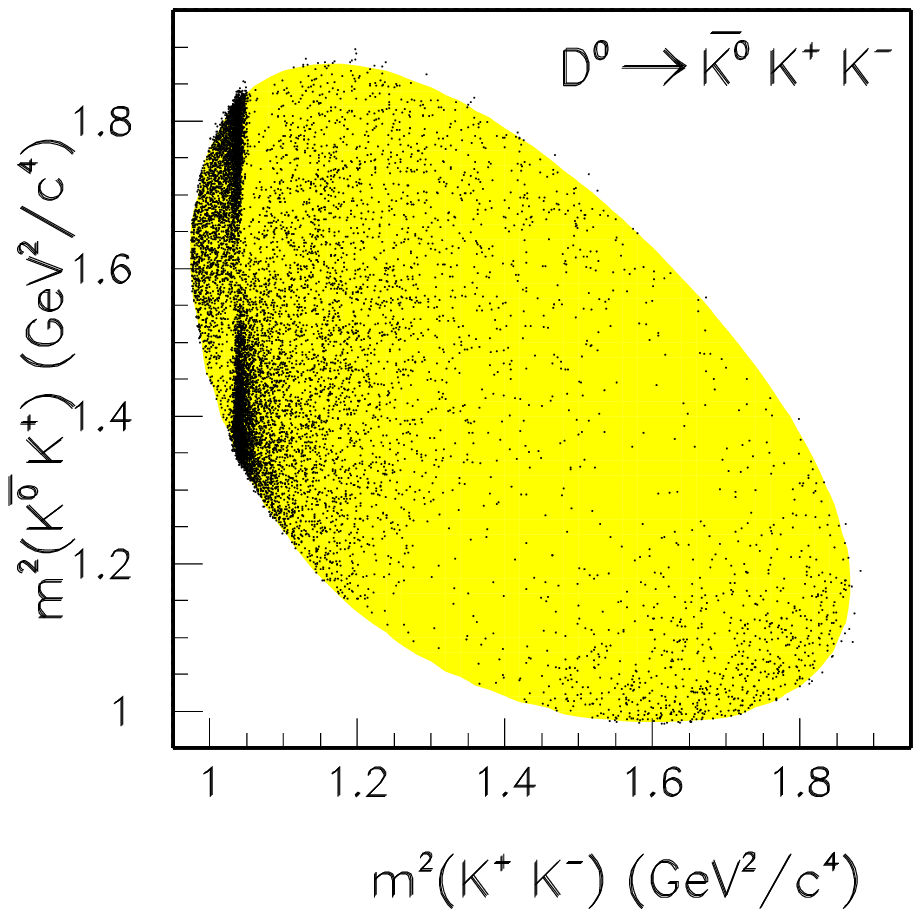}
\caption{Dalitz plot of $D^0 \to \overline{K}{}^0 K^+ K^-$.
} 
\label{fig:fig3}
\end{center}
\end{figure}

Selecting events within $\pm 2 \sigma$ of the fitted $D^0$ mass value,
a signal fraction of 97.3\% is obtained for the 12540 events selected. 
The Dalitz plot for these $D^0 \to \overline{K}{}^0 K^+ K^-$ candidates 
is shown 
in Fig.~\ref{fig:fig3}.
In the $K^+ K^-$ threshold region, a strong $\phi(1020)$ signal is observed, 
together with a rather broad structure. 
A large asymmetry with respect to the $\overline{K}{}^0 K^+$ axis can also 
be seen 
in the vicinity of the $\phi(1020)$ signal, which is most probably the 
result of interference between  
$S$ and $P$-wave amplitude contributions to the $K^+ K^-$ system.
The $f_0(980)$ and $a_0(980)$ $S$-wave resonances are, in fact, just below the 
$K^+ K^-$ 
threshold, and might be expected to contribute in the vicinity of $\phi(1020)$.
An accumulation of 
events due to a charged $a_0(980)^+$ can be observed on the lower right edge 
of the Dalitz plot. This contribution, however, does not overlap with 
the $\phi(1020)$ region and 
this allows the $K^+K^-$ scalar and vector components to be separated using a 
partial wave analysis in the low mass $K^+K^-$ region.

\section{Partial Wave Analysis}
\begin{figure}
\begin{center}
\includegraphics[width=8cm]{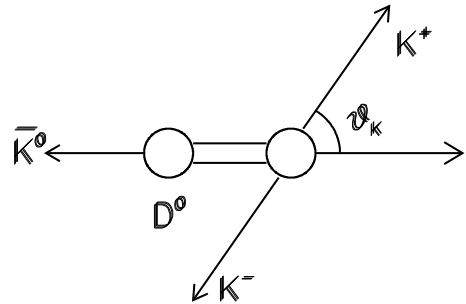}
\caption{The kinematics describing the production of the $K^+ K^-$ system
in the threshold region.
} 
\label{fig:fig4}
\end{center}
\end{figure}

\begin{figure*}
\begin{center}
\includegraphics[width=14cm]{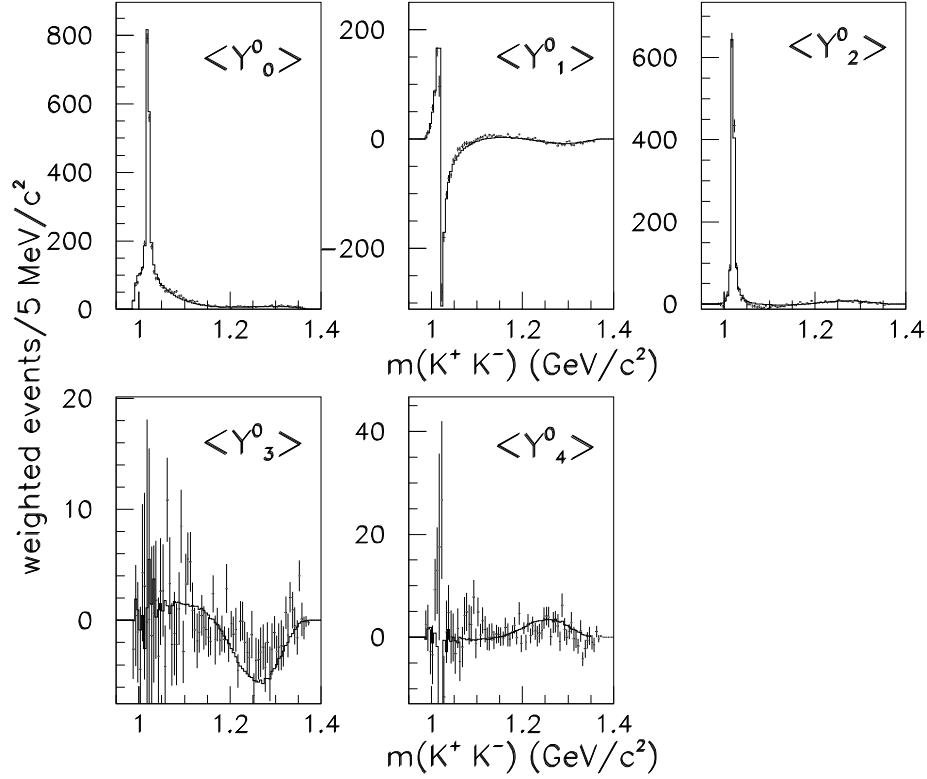}
\caption{The unnormalized spherical harmonic moments $\left <Y^0_L \right >$ 
as functions of $K^+ K^-$
invariant mass. The histograms represent the result of the full Dalitz plot
analysis. 
} 
\label{fig:fig5}
\end{center}
\end{figure*}

It is assumed that near threshold the production of the $K^+K^-$ 
system can be described in terms of the diagram shown in
Fig.~\ref{fig:fig4}.
The helicity angle, $\theta_K$, is then defined as the 
angle between the $K^+$ for $D^0$ (or $K^-$ for $\overline{D}{}^0$)
in the $K^+ K^-$ rest frame and 
the $K^+ K^-$ direction in the $D^0$ (or $\overline{K}{}^0$)
rest frame. The $K^+ K^-$ mass distribution has been modified by weighting 
each $D^0$ candidate by the spherical harmonic $Y_L^0(\cos \theta_K)$ (L=0-4)
divided by its (Dalitz-plot-dependent) fitted efficiency.
The resulting distributions $\left<Y^0_L \right>$ are shown in 
Fig.~\ref{fig:fig5} and are proportional to the $K^+K^-$ mass-dependent
harmonic moments.
It is found that all the $\left<Y^0_L \right>$ moments are
small or consistent with zero, except for 
$\left<Y^0_0 \right>$, $\left<Y^0_1 \right>$ and $\left<Y^0_2 \right>$.
  
In order to interpret these distributions a simple 
partial wave analysis has been performed, involving only $S$- and
$P$-wave amplitudes. This results in the following set of equations~\cite{chung}:

$$\sqrt{4 \pi} \left<Y^0_0 \right> = S^2 + P^2$$
$$\sqrt{4 \pi} \left<Y^0_1 \right> = 2 \mid S \mid \mid P \mid \cos \phi_{SP} \qquad (3)$$
$$\sqrt{4 \pi} \left<Y^0_2 \right> = \frac{2}{\sqrt 5} P^2,$$

\noindent
where $S$ and $P$ are proportional to the size of the $S$- and $P$-wave
contributions and $\phi_{SP}$ is their relative phase.
Under these assumptions, the $\left<Y^0_2 \right>$ moment is proportional to $P^2$ 
so that it is natural that the $\phi(1020)$ appears free of background, 
as is observed.
This distribution
has been fit using the following relativistic $P$-wave Breit-Wigner.

For a resonance $r \to AB$,  $BW(m)$ is written as 
$$BW(m) = \frac{F_r}{m_r^2 - m_{AB}^2 - i \Gamma_{AB}m_r} \qquad (4)$$
where for a spin J=1 particle $F_r$ is the Blatt-Weisskopf damping 
factor~\cite{dump} 
$$F_r = \frac{\sqrt{1+(R q_r)^2}}{\sqrt{1+(R q_{AB})^2}}$$
and $R$ has been fixed to $R=1.5$ ${\rm GeV}^{-1}$.
In Eq. (4):
$$\Gamma_{AB} = \Gamma_r (\frac{q_{AB}}{q_r})^{2J+1} (\frac{m_r}{m_{AB}})F^2_r$$
where 
$q_{AB}$ ($q_r$) is the momentum of either daughter
in the $AB$ ($r$) rest frame.

\begin{figure}
\begin{center}
\includegraphics[width=9cm]{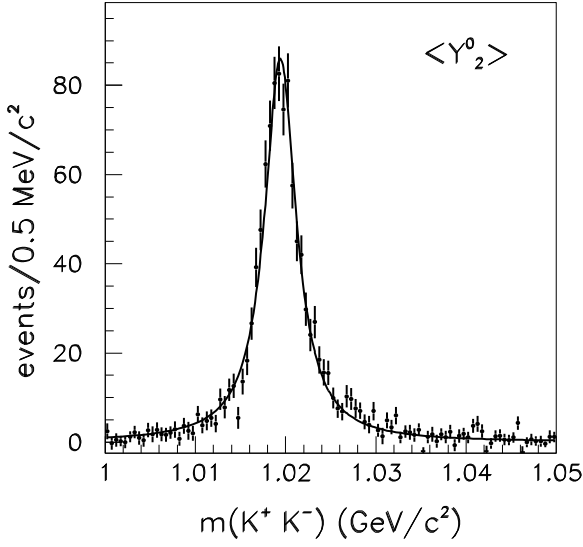}
\caption{$\left<Y^0_2 \right>$ spherical harmonic moment as a function of the $K^+ K^-$
effective mass. The line is the result from the fit with a relativistic 
spin-1 Breit Wigner. 
} 
\label{fig:fig6}
\end{center}
\end{figure}

The fit yields
the
following parameters:
\begin{center}
$m_\phi$ = 1019.63 $\pm$ 0.07, $\Gamma_\phi$ = 4.28 $\pm$ 0.13 MeV/$c^2$
\end{center}
in agreement with PDG values (statistical errors only).
The fit is shown in Fig.~\ref{fig:fig6}.

A strong $S-P$ interference is evidenced by the rapid motion
of the $\left<Y^0_1 \right>$ moment in Fig.~\ref{fig:fig5} in the $\phi(1020)$ mass region.

The above system of equations (3) can be solved directly for $S^2$, $P^2$ and
$\cos \phi_{SP}$. However, since these amplitudes are defined in a $D^0$ decay,
it is necessary to correct for phase space. 
This has been achieved by using the $K^+ K^-$ and $\overline{K}{}^0 K^+$ 
mass spectra obtained from the Monte Carlo generation of $D^0$ decays 
to $\overline{K}{}^0 K^+ K^-$ according to phase space.
The $D^0$ mass distribution has been generated in this Monte Carlo as a Gaussian having the
experimental values of mass and mass resolution.

\begin{figure*}
\begin{center}
\includegraphics[width=14cm]{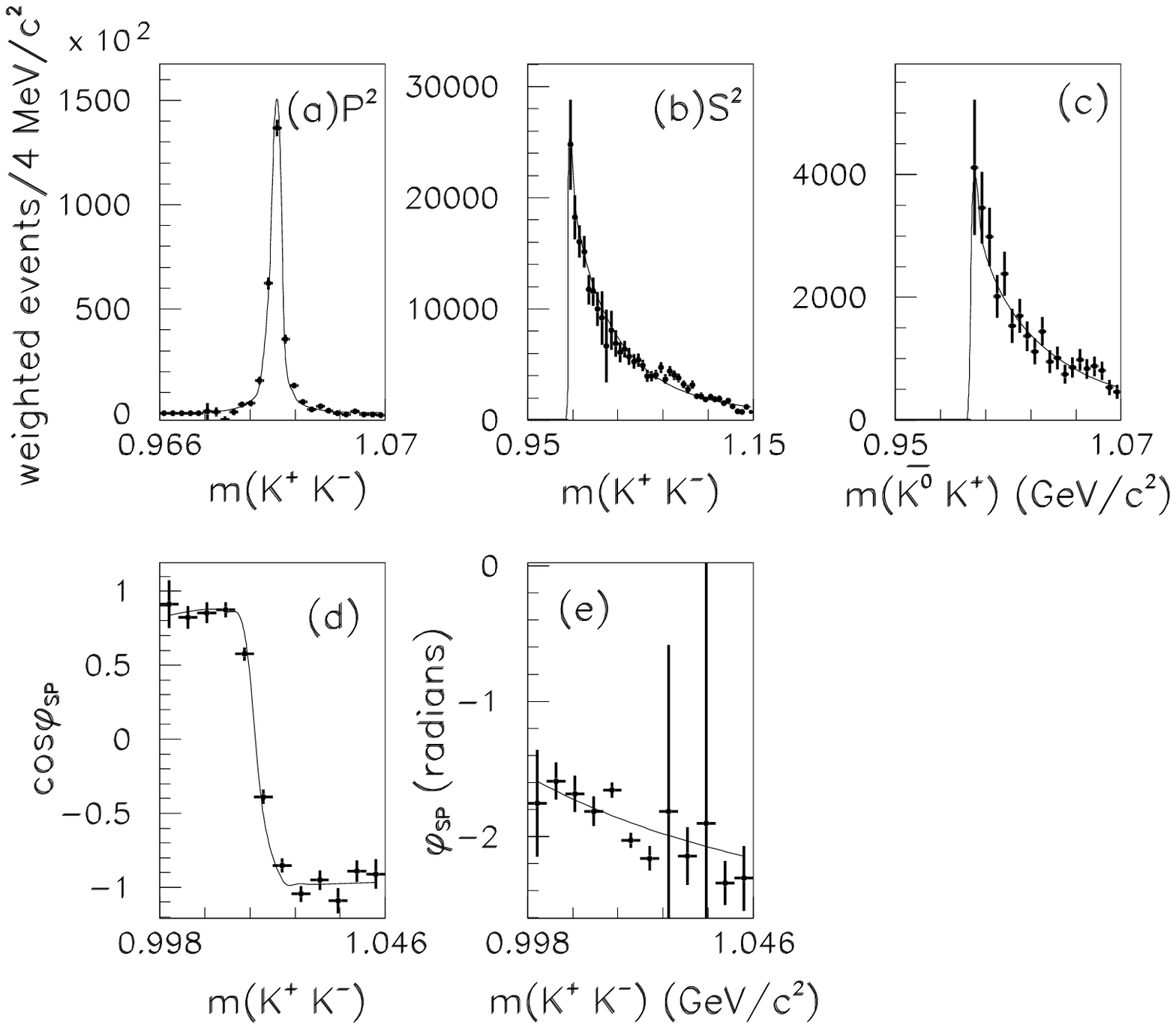}
\caption{Results from the $K^+ K^-$ Partial Wave Analysis corrected 
for phase space. (a) $P$-wave strength, (b) $S$-wave strength.
(c) $m(\overline{K}{}^0 K^+)$ distribution, (d) $\cos\phi_{SP}$
in the $\phi(1020)$ region. (e) $\phi_{SP}$ in the threshold
region
after having subtracted the fitted $\phi(1020)$ phase motion 
shown in (d).
The lines correspond to the fit described in the text.
} 
\label{fig:fig7}
\end{center}
\end{figure*}

The phase space corrected spectra are shown in Fig.~\ref{fig:fig7}.

The distributions have been fitted using the following model:
\begin{itemize}
\item{} The P-wave is entirely due to the $\phi(1020)$ meson (Fig.~\ref{fig:fig7}(a)). 
\item{} The scalar contribution in the $K^+ K^-$ mass projection is 
entirely due to the $a_0(980)^0$ (Fig.~\ref{fig:fig7}(b)). 
\item{} The $\overline{K}{}^0 K^+$ mass distribution is entirely due to 
$a_0(980)^+$ (Fig.~\ref{fig:fig7}(c)).
\item{} The angle $\phi_{SP}$ (Fig.~\ref{fig:fig7}(d)) is obtained fitting the
S, P waves and $\cos \phi_{SP}$ with $c_{a_0}BW_{a_0} + c_{\phi}BW_{\phi}e^{i \alpha}$.
Here $BW_{a_0}$ and $BW_{\phi}$ are the Breit-Wigner describing the $a_0(980)$ 
and $\phi(1020)$ resonances.
\end{itemize}
The $a_0(980)$ scalar resonance has a mass very close
to the $\bar K K$ threshold and decays mostly to $\eta \pi$. It has been
described by a coupled channel Breit Wigner of the form:
$$BW_{ch}(a_0)(m) = \frac {g_{\bar K K}}{m^2_0 - m^2 - 
i(\rho_{\eta \pi} g_{\eta \pi}^2 + \rho_{\bar K K} g_{\bar K K}^2)} \quad (5)$$
\noindent
where $\rho(m) = 2 q/m$ while $g_{\eta \pi}$ and $g_{\bar K K}$ describe
the $a_0(980)$ couplings to the $\eta \pi$ and $\bar K K$ systems respectively.

The best measurements of the $a_0(980)$ parameters come from 
the Crystal Barrel experiment~\cite{cbar}, 
in $\bar p p$ annihilations, and are the following:
\begin{center}
$m_0$=999$\pm$2 MeV/$c^2$, $g_{\eta \pi}$=324$\pm$15 (MeV)$^{1/2}$, 
$\frac{g^2_{\eta \pi}}{g^2_{\bar K K}}$=1.03$\pm$0.14.
\end{center}
This corresponds to a value of $g_{\bar K K}=329 \pm 27$ (MeV)$^{1/2}$.

Since in the current analysis only the $\bar K K$ projections are available, 
it is not possible to measure $m_0$ and $g_{\eta \pi}$. 
Therefore, these two quantities 
have been fixed to the 
Crystal Barrel measurements. 
The parameter $g_{\bar K K}$, on the other hand,
has been left free in the 
fit. 
The result is (statistical error only):
\begin{center}
$g_{\bar K K}$ = 464 $\pm$ 29 (MeV)$^{1/2}.$
\end{center}
Figure~\ref{fig:fig7}(e) shows the residual $a_0(980)$ phase, obtained
by first computing $\phi_{SP}$ in the range (0,$\pi$) and then subtracting the known phase
motion due to the $\phi(1020)$ resonance. The fit gives 
a value of a relative phase $\alpha=2.12 \pm 0.04$ and has a $\chi^2/NDF$=167/92.
The fit is of rather poor quality, indicating an undetermined source of
systematic uncertainty comparable with the statistical uncertainty. However 
the issue related to the determination of $g_{\bar K K}$ will be rediscussed 
in the complete Dalitz plot analysis described in section VIII.  

The entire procedure has been tested with Monte Carlo simulations with
different input values of the $a_0(980)$ parameters. The partial wave 
analysis performed on these simulated data yielded the input value of 
$g_{\bar K K}$, within the errors. 

In this fit the possible presence of an $f_0(980)$
contribution has not been considered.
This assumption can be tested by comparing the $K^+ K^-$ and 
$\overline{K}{}^0 K^+$
phase space corrected mass distributions. Since the $f_0(980)$ has isospin 0, it cannot 
decay to $\overline{K}{}^0 K^+$. Therefore an excess in the
$K^+ K^-$ mass spectrum with respect to $\overline{K}{}^0 K^+$ would indicate
the presence of an $f_0(980)$ contribution. 

\begin{figure}
\begin{center}
\includegraphics[width=9cm]{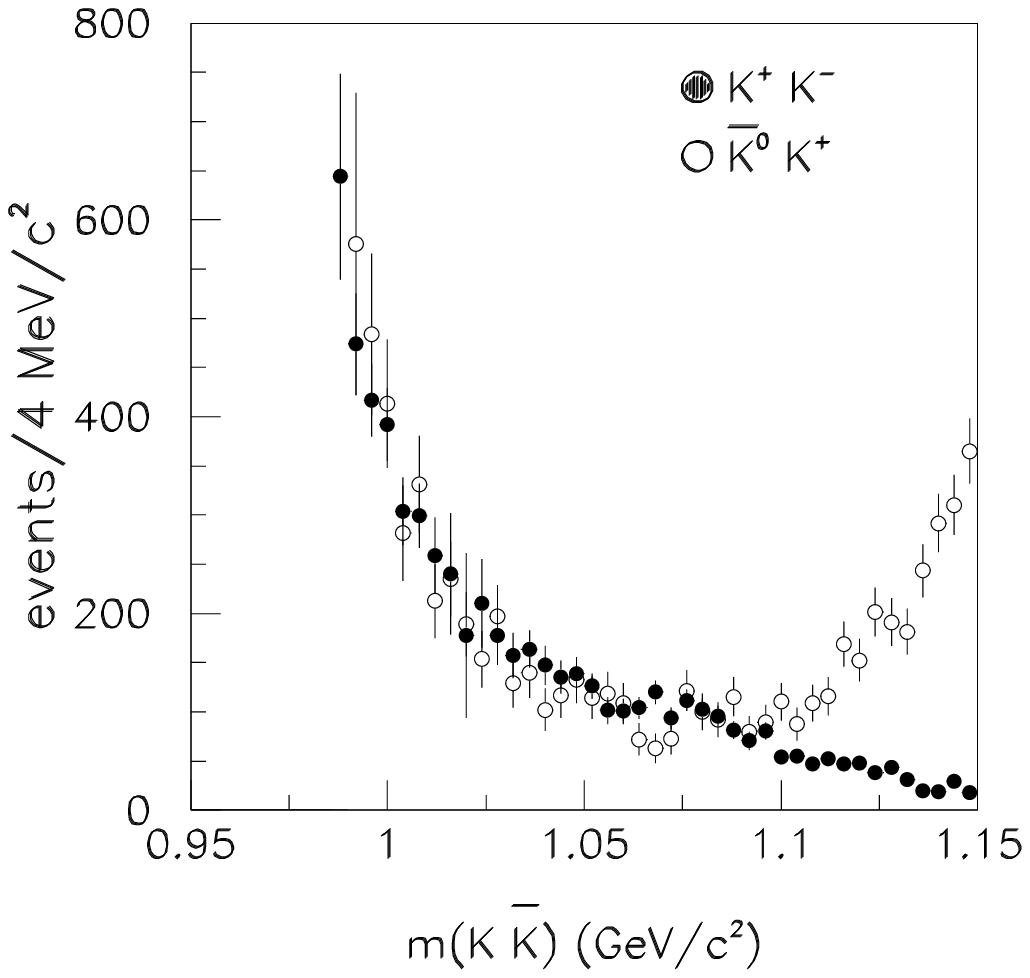}
\caption{Comparison between the phase-space-corrected $K^+ K^-$ and 
$\overline{K}{}^0 K^+$
normalised to the same
area in the mass region between 0.992 and 1.05 GeV/$c^2$.
} 
\label{fig:fig8}
\end{center}
\end{figure}

Figure~\ref{fig:fig8} compares the $K^+ K^-$ and 
$\overline{K}{}^0 K^+$ 
mass distributions, normalised to the same area between 0.992 and 1.05 
GeV/$c^2$ and corrected for phase space. It is possible to observe that 
the two 
distributions show a good agreement, supporting the argument that the 
$f_0(980)$ contribution is small. Notice that the enhanced 
$\overline{K}{}^0 K^+$ signal level above 1.1 GeV/$c^2$ is the result 
of the $\phi(1020)$
reflection.

The resulting scalar components of the $K^+ K^-$ and $\overline{K}{}^0 K^+$ mass
distributions, corrected for phase space, are tabulated as a function 
of mass in Table~\ref{tab:one}.
\begin{table}[tbp]
\caption{$K^+ K^-$ and $\overline{K}{}^0 K^+$ scalar mass projections corrected for
phase space in arbitrary units.}
\label{tab:one}
\begin{center}
\vskip -0.2cm
\begin{tabular}{lcc}
\hline

mass (GeV/$c^2$) & $K^+ K^-$  & $\overline{K}{}^0 K^+$ \cr

\hline
0.988 & 644 $\pm$  105  &  \cr
0.992 & 474 $\pm$ 52 & 575 $\pm$ 154 \cr
0.996 & 417 $\pm$ 37 & 484 $\pm$ 82 \cr 
1.000  & 392 $\pm$ 37 & 414 $\pm$ 65 \cr 
1.004 & 304 $\pm$ 35 & 282 $\pm$ 48 \cr 
1.008 & 299 $\pm$ 33 & 331 $\pm$ 49 \cr 
1.012 & 259 $\pm$ 39 & 213 $\pm$ 38 \cr
1.016 & 240 $\pm$ 62 & 235 $\pm$ 38 \cr
1.020  & 178 $\pm$ 84 & 189 $\pm$ 33 \cr
1.024 & 210 $\pm$ 45 & 153 $\pm$ 28 \cr
1.028 & 178 $\pm$ 30 & 197 $\pm$ 32 \cr
1.032 & 157 $\pm$ 23 & 129 $\pm$ 25 \cr
1.036 & 164 $\pm$ 19 &  140 $\pm$ 25 \cr
1.040 & 147 $\pm$ 20 & 102 $\pm$ 21 \cr
1.044 & 135 $\pm$ 17 &117 $\pm$ 22 \cr
1.048 & 139 $\pm$ 15 & 132 $\pm$ 23 \cr
1.052 & 126 $\pm$ 13 & 114 $\pm$ 22 \cr
1.056 & 101 $\pm$ 14 & 119 $\pm$ 22 \cr
1.060 & 101 $\pm$ 12 & 108 $\pm$ 20 \cr
1.064 & 104 $\pm$ 11 & 72  $\pm$  17 \cr
1.068 & 120 $\pm$ 12 & 63 $\pm$ 15 \cr
\hline
\end{tabular}
\end{center}
\end{table} 

\begin{figure}
\begin{center}
\includegraphics[width=8cm]{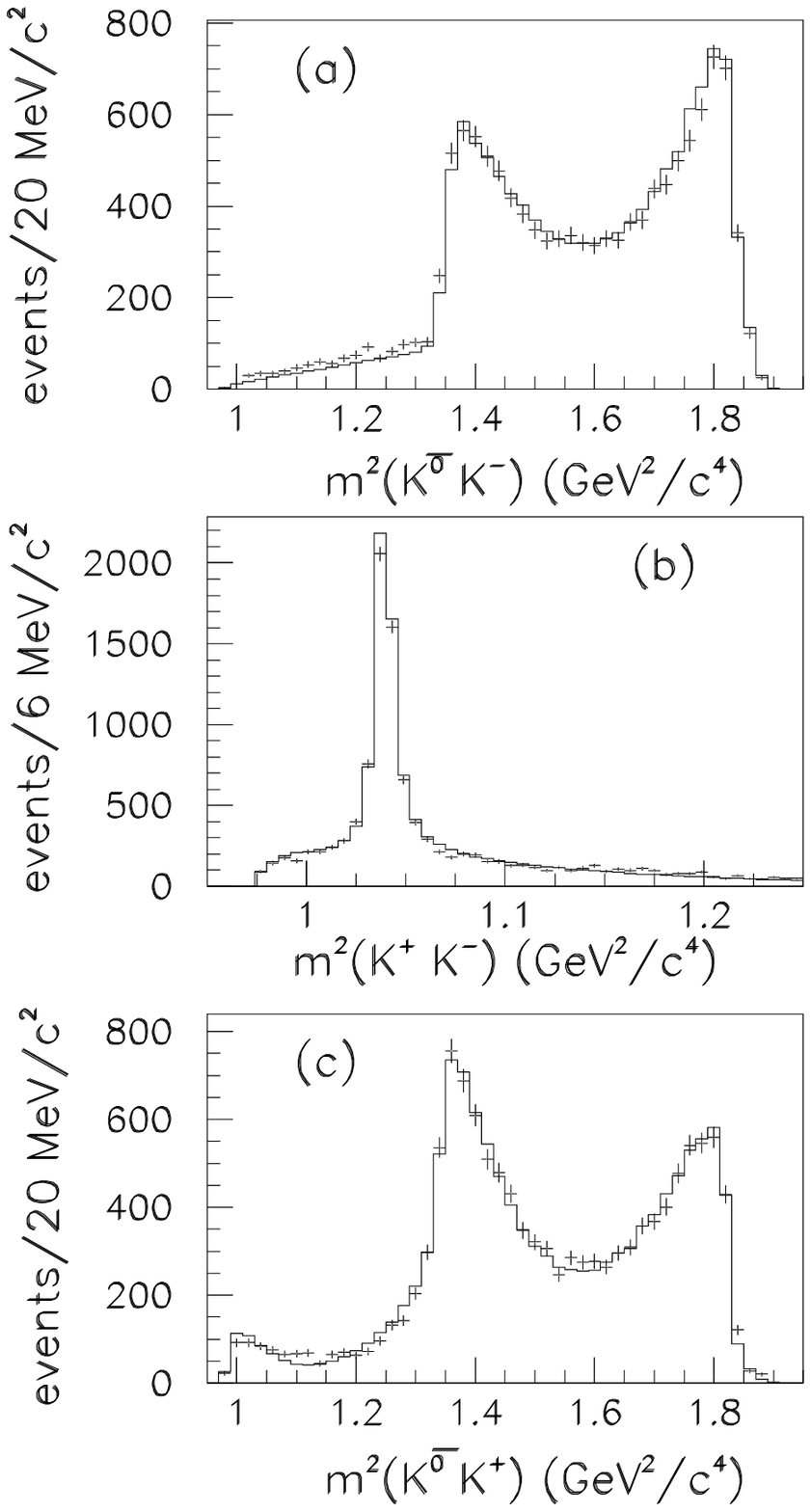}
\caption{Dalitz plot projections for  $D^0 \to \overline{K}{}^0 K^+ K^-$. 
The data are
represented with error bars; the histogram is 
the projection of the fit described in the text.}
\label{fig:fig9}
\end{center}
\end{figure} 

\section{Dalitz Plot Analysis of $D^0 \to \overline{K}{}^0 K^+ K^-$.}

An unbinned maximum likelihood fit has been performed
for the decay
$D^0 \to \overline{K}{}^0 K^+ K^-$
in order to use the distribution of events 
in the Dalitz plot to determine the relative amplitudes and phases 
of intermediate resonant and non-resonant states.
 
The likelihood function has been written
in the following way:

$$L = \beta \cdot G(m) \epsilon(m_x^2,m_y^2) \frac {\sum_{i,j} c_i c_j^* A_i A_j^*}
{\int{\sum_{i,j} c_i c_j^*A_i A_j^* \epsilon(m_x^2,m_y^2)} dm_x^2 dm_y^2}$$
$$ + (1 - \beta) \qquad (6)$$
In this expression, $\beta$ represents the fraction of signal obtained from the fit to
the mass spectrum and $\epsilon(m_x^2,m_y^2)$ is the fitted efficiency on the
Dalitz plot. The Gaussian function $G(m)$ describes the $D^0$ line shape 
normalised within the
$\pm 2 \sigma$ cutoff used to perform the Dalitz plot analysis. 
It is assumed that the background events, described by the second term
in Eq.~(6), uniformly populate the Dalitz plot.
This assumption has been 
verified by examining events in the $D^0$ side bands.
The output from the fit is the set of
complex coefficients $c_i$.

In Eq.~(6), the integrals have been computed using Monte Carlo events
while taking into
account the efficiency on the Dalitz plot.
The branching fraction for the resonant or non-resonant contribution $i$
is defined by the following expression:
$$f_i = \frac {|c_i|^2 \int |A_i|^2 dm_x^2 dm_y^2}
{\sum_{j,k} c_j c_k^* \int A_j A_k^* dm_x^2 dm_y^2}. $$
The fractions $f_i$ do not necessarily add up to 1 because of interference
effects among the amplitudes. The errors on the 
fractions have been evaluated by propagating the 
full covariance matrix obtained
from the fit.

The phase of each amplitude is measured with respect to 
$\overline{K}{}^0 a_0(980)^0$ which gives
the largest contribution. 
The amplitudes $A_i$ are represented by the product of complex 
Breit-Wigner $BW(m)$ (Eq.~(4)) and angular terms $T (\Omega)$~\cite{cleo}:
$$ A = BW(m) \times T (\Omega). $$
The $f_0(980)$ resonance has been described using a coupled 
channel Breit-Wigner function with parameters taken
from the WA76~\cite{wa76}, E791~\cite{e791}, and BES~\cite{bes}.
The $a_0(980)$ has been parametrized using the results from
the partial wave analysis discussed above.

The parameters of the $\phi(1020)$ meson have been fixed to the values
obtained from the fit to the $\left<Y^0_2 \right>$ moment described earlier.
The non-resonant contribution (NR) is represented by a constant
term with a free phase.

Systematic errors on the fitted fractions have been evaluated by making 
different assumptions in the fits.
For example, in one test, the efficiency on the Dalitz plot has been set to a 
constant value. In other tests the resonance parameters of $f_0(980)$,
$a_0(980)$ and $f_0(1400)$ have been fixed to values 
obtained from a variety of experiments. 

The doubly-Cabibbo-suppressed contribution
(DCS) $K^+ a_0(980)^-$, whose presence should appear like an 
$a_0(980)^-$ in the wrong sign combination $\overline{K}{}^0 K^-$, has been 
also included in the fit.

\section{Results from the Dalitz plot analysis.}

The $D^0 \to \overline{K}{}^0 K^+ K^-$ Dalitz plot projections 
together with the fit
results are shown in Fig.~\ref{fig:fig9}. 

Figure~\ref{fig:fig7} shows 
the fit projections onto the $\left<Y^0_L \right>$ moments.
The fit produces a reasonable representation of the data for all of the 
projections.
The $\chi^2$ computed on the Dalitz plot gives a value of 
$\chi^2/NDF$=983/774. The sum of the fractions is $130.7 \pm 2.2 \pm 8.4$\%.
The regions of higher $\chi^2$ are distributed rather uniformly on the Dalitz
plot. Attempts to improve the fit quality by including other contributions
did not give better results. One particular problem
found in these fits is that including too many scalar amplitudes caused the 
fit to diverge, producing a sum of fractions well above 200\% along with small
improvements of the fit quality.

The final fit results showing fractions, amplitudes and phases
are summarised in Table~\ref{tab:res}. 
For $\overline{K}{}^0 f_0(980)$ and $K^+ a_0(980)^-$ (DCS), being
consistent
with zero,  only the fractions have 
been tabulated.

\begin{table*}[tbp]
\caption{Results from the Dalitz plot analysis of $D^0 \to \overline{K}{}^0 K^+ K^-$. The fits have been performed using the value of $g_{\bar K K} = 464$ (MeV)$^{1/2}$ resulting from the partial wave analysis.}
\label{tab:res}
\begin{center}
\vskip -0.2cm
\begin{tabular}{lllll}
\hline
Final state & Amplitude & Phase (radians) & Fraction (\%) \cr
\hline
$\overline{K}{}^0 a_0(980)^0$ & 1. & 0. & 66.4 $\pm$ 1.6 $\pm$ 7.0 \cr
\hline
$\overline{K}{}^0 \phi(1020)$ & 0.437 $\pm$ 0.006 $\pm$ 0.060 & 1.91 $\pm$ 0.02 $\pm$
 0.10 &  45.9 $\pm$ 0.7 $\pm$ 0.7 \cr
\hline 
$K^- a_0(980)^+$ & 0.460 $\pm$ 0.017 $\pm$ 0.056 & 3.59 $\pm$ 0.05 $\pm$ 0.20 &
 13.4 $\pm$ 1.1 $\pm$ 3.7 \cr
\hline 
$\overline{K}{}^0 f_0(1400)$ & 0.435 $\pm$ 0.033 $\pm$ 0.162 & -2.63 $\pm$ 0.10 $\pm$ 0
.71 & 3.8 $\pm$ 0.7 $\pm$ 2.3 \cr
\hline
$\overline{K}{}^0 f_0(980)$ & & &  0.4 $\pm$ 0.2 $\pm$ 0.8 \cr 
\hline 
$K^+ a_0(980)^-$ & & &  0.8 $\pm$ 0.3 $\pm$ 0.8 \cr
\hline
Sum & & & 130.7 $\pm$ 2.2 $\pm$ 8.4 \cr
\hline
\end{tabular}
\end{center}
\end{table*} 

The results from the Dalitz plot analysis can be summarised as follows:
\begin{itemize}
\item{} The decay is dominated by $D^0 \to \overline{K}{}^0 a_0(980)^0$, 
$D^0 \to \overline{K}{}^0 \phi(1020)$ and $D^0 \to K^- a_0(980)^+$. 
\item{} The $f_0(980)$ contribution is consistent with zero, 
even after assuming various $f_0(980)$ lineshape parameters~\cite{wa76}~\cite{e791}~\cite{bes}.
\item{} The DCS contribution is consistent with zero, regardless of the 
$a_0(980)$ parametrisation.
\item{} The remaining contribution is not consistent with being 
uniform, but can be described by the tail of a broad resonance, for example
the $f_0(1400)$ which peaks well outside the phase space.
It is not possible to derive its parameters from our data, but several 
parametrizations have been tried, in particular those 
from $J/\psi$ decays~\cite{bes} and from 
$D^+_s \to \pi^+ \pi^+ \pi^-$~\cite{e791} getting in all cases improved fits. 
\item{} In one of the fits the $f_0(1400)$ contribution has been replaced 
by a non-resonant 
contribution, obtaining a fraction of $2.5 \pm 0.5$\%. However the likelihood 
value for this fit  was worse $\Delta (2 \log L)=56$.
\end{itemize} 
For the $\overline{K}{}^0 f_0(980)$ and DCS contributions upper limits have been
 computed.
Combining statistical and systematic errors in quadrature, the 
following 95 \% c.l. upper limits on the fractions have been obtained:
$$BF(D^0 \to \overline{K}{}^0 f_0(980)(\to K^+ K^-)) < 2.1\%$$
$$BF(D^0 \to K^+ a_0(980)^-(\to \overline{K}{}^0 K^-)) (DCS) < 2.5\%.$$

A test has been performed by leaving $g_{\bar K K}$ as a free parameter 
in the Dalitz plot analysis. In this test the other parameters describing
the $a_0(980)$ ($m_0$ and $g_{\eta \pi}$) have been allowed to vary within 
their measurement errors from the Crystal Barrel experiment.
The resulting central value of $g_{\bar K K}$ is
473 (MeV)$^{1/2}$ with a maximum deviation of 39 (MeV)$^{1/2}$, in good
agreement with the value obtained using the partial wave analysis. 
The difference between
the values, added in quadrature with the above maximum deviation,
has been taken as an estimate of the systematic error: 
$$g_{\bar K K} = 473 \pm 29 \;(\text{stat.}) \pm 40 \;(\text{syst.}) ({\rm MeV})^{1/2}.$$
This value differs significantly from the Crystal Barrel measurement. An improvement 
of this measurement can be foreseen by adding data from the $a_0(980) \to \eta \pi$ decay 
mode such as $D^0 \to K^0_s \eta \pi^0$. This $D^0$ decay mode
has been studied by the CLEO~\cite{cleo1} experiment (with rather limited statistics)
finding a $D^0 \to  \overline{K}{}^0 a_0(980)^0$ dominant contribution. 

\begin{figure}
\begin{center}
\includegraphics[width=9cm]{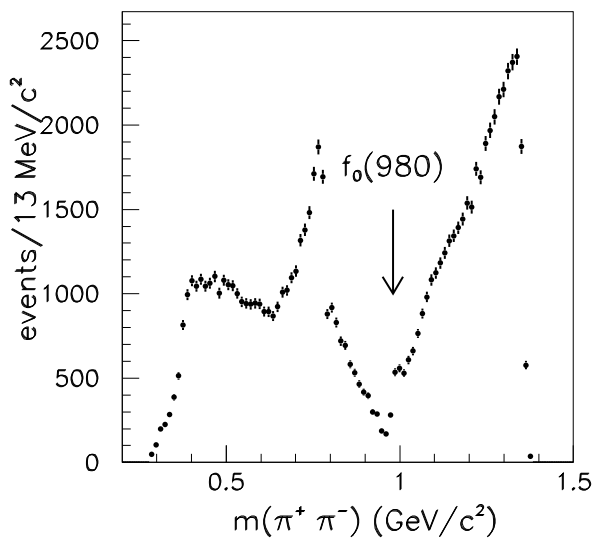}
\caption{$\pi^+ \pi^-$ effective mass from $D^0 \to \overline{K}{}^0 \pi^+ \pi^-$. The
arrow indicates the position of the $f_0(980)$.
} 
\label{fig:fig10}
\end{center}
\end{figure}

A large uncertainty is included in the upper 
limit on the presence of $f_0(980)$ in this $D^0$ decay mode 
due to the poor knowledge of the $f_0(980)$ 
parameters. A small signal of $f_0(980)$ is indeed present (in this case as a shoulder) in the 
$D^0 \to \overline{K}{}^0 \pi^+ \pi^-$ as shown in Fig.~\ref{fig:fig10}.  

Dalitz plot analyses 
of this $D^0$ decay channel have been performed by \babar~\cite{gamma1} and Belle~\cite{gamma2}
finding ($\approx$ 5.5\%) as decay fraction for $D^0 \to \overline{K}{}^0 f_0(980)$.
However, a reliable estimate of the expected contribution of the
$f_0(980)$ in $D^0 \to \overline{K}{}^0 K^+ K^-$ decay is not possible until more accurate
measurements of the $f_0(980)$ parameters and couplings become available.
This can be performed, for example, by using high 
statistics samples of $D^+_s \to \bar K K \pi^+$ and $D^+_s \to \pi^+ \pi^+ \pi^-$ decays.

\begin{table*}[tb]
\caption{Results from the Dalitz plot Analysis of
 $D^0 \to \overline{K}{}^0 K^+ K^-$ separated for $D^0$ and $\overline{D}{}^0$.}
\label{tab:res1}
\begin{center}
\vskip -0.2cm
\begin{tabular}{lcccc}
\hline
Decay mode & fraction (\%) & amplitude & phase (radians) & $\chi^2/NDF$ \cr
\hline
$D^0 \to \overline{K}{}^0 a_0(980)^0$ &  66.5 $\pm$ 2.0 & 1. & 0. & 671/649  \cr 
$\overline{D}{}^0\to K^0 a_0(980)^0$ &  66.3 $\pm$ 2.0 & 1. & 0. & 643/646 \cr 
\hline
$D^0 \to \overline{K}{}^0 \phi(1020)$ & 46.3 $\pm$ 0.8 &  0.438 $\pm$ 0.009 &  1.93 $\pm$  0.03 &\cr  
$\overline{D}{}^0 \to K^0 \phi(1020)$ & 45.6 $\pm$ 0.8 &  0.435 $\pm$ 0.009 &  1.88 $\pm$  0.03 &\cr 
\hline 
$D^0 \to K^- a_0(980)^+$ & 13.2 $\pm$ 1.3 & 0.456 $\pm$ 0.025 & 3.58 $\pm$ 0.07 & \cr 
$\overline{D}{}^0 \to K^+ a_0(980)^-$ & 13.6 $\pm$ 1.3 & 0.463 $\pm$ 0.025 & 3.59 $\pm$ 0.07 & \cr 
\hline 
$D^0 \to \overline{K}{}^0 f_0(1400)$ & 4.1 $\pm$ 0.9 & 0.451 $\pm$ 0.047 & -2.58 $\pm$ 0.13 & \cr
$\overline{D}{}^0 \to K^0 f_0(1400)$ & 3.6 $\pm$ 0.9 & 0.421 $\pm$ 0.038 & -2.68 $\pm$ 0.14 & \cr
\hline 
\end{tabular}
\end{center}
\end{table*}

\section{Search for CP asymmetries on the Dalitz plot.}

A search for CP asymmetries on the Dalitz plot has been performed.
Table~\ref{tab:res1} shows the
results from the Dalitz plot analysis performed separately
for $D^0$ and $\overline{D}{}^0$. Notice that in these two fits
good values of $\chi^2/NDF$ have been obtained.

We do not observe any statistically significant asymmetries in fractions, 
amplitudes, 
or phases between $D^0$ and $\overline{D}{}^0$.

\section{Summary}
\label{sec:Summary}

A Dalitz plot analysis of the $D^0$ hadronic 
decay 
$D^0 \to \overline{K}{}^0 K^+ K^-$ has been performed.
The following ratio of branching fractions has been obtained: 
$$BR = \frac{\Gamma(D^0 \to \overline{K}{}^0 K^+ K^-)}{\Gamma(D^0 \to \overline{K}{}^0 \pi^+ \pi^-)} =$$
$$  (15.8 \pm 0.1 \;(\text{stat.})
\pm 0.5 \;(\text{syst.})) \times 10^{-2}.$$
The Dalitz plot analysis indicates that the channel is dominated by $D^0 \to \overline{K}{}^0 a_0(980)^0$, 
$D^0 \to \overline{K}{}^0 \phi(1020)$ and $D^0 \to K^- a_0(980)^+$.  
The $a_0(980) \to \bar K K$ lineshape has been extracted with little 
background. 

The Dalitz plot analysis of $D^0$ and $\overline{D}{}^0$ do not show any statistically
significant asymmetries in fractions, amplitudes, or phases.

\section{Acknowledgments}
\label{sec:Acknowledgments}

We are grateful for the 
extraordinary contributions of our \pep2\ colleagues in
achieving the excellent luminosity and machine conditions
that have made this work possible.
The success of this project also relies critically on the 
expertise and dedication of the computing organizations that 
support \babar.
The collaborating institutions wish to thank 
SLAC for its support and the kind hospitality extended to them. 
This work is supported by the
US Department of Energy
and National Science Foundation, the
Natural Sciences and Engineering Research Council (Canada),
Institute of High Energy Physics (China), the
Commissariat \`a l'Energie Atomique and
Institut National de Physique Nucl\'eaire et de Physique des Particules
(France), the
Bundesministerium f\"ur Bildung und Forschung and
Deutsche Forschungsgemeinschaft
(Germany), the
Istituto Nazionale di Fisica Nucleare (Italy),
the Foundation for Fundamental Research on Matter (The Netherlands),
the Research Council of Norway, the
Ministry of Science and Technology of the Russian Federation, and the
Particle Physics and Astronomy Research Council (United Kingdom). 
Individuals have received support from 
CONACyT (Mexico),
the A. P. Sloan Foundation, 
the Research Corporation,
and the Alexander von Humboldt Foundation.

\end{document}